\documentclass[aip,reprint,amsmath,amssymb,superscriptaddress]{revtex4-2}

\usepackage{verbatim} 

\usepackage{graphicx}
\usepackage{dcolumn}

\usepackage{amstext,amsmath,amssymb,amsfonts}
\usepackage{bm}

\usepackage[utf8]{inputenc}
\usepackage[T1]{fontenc}
\usepackage{txfonts}
\usepackage{mathtools}
\DeclareMathAlphabet{\mathpzc}{OT1}{pzc}{m}{it}
\DeclareFontFamily{OT1}{pzc}{}
\DeclareFontShape{OT1}{pzc}{m}{it}{<-> s * [1.100] pzcmi7t}{}
\DeclareMathAlphabet{\mathpzc}{OT1}{pzc}{m}{it}

\def\app#1#2{%
  \mathrel{%
    \setbox0=\hbox{$#1\sim$}%
    \setbox2=\hbox{%
      \rlap{\hbox{$#1\propto$}}%
      \lower1.1\ht0\box0%
    }%
    \raise0.25\ht2\box2%
  }%
}

\usepackage{hyperref}

\usepackage{color}
\definecolor{lightblue}{rgb}{0.2,0.2,0.7}
\definecolor{darkblue}{rgb}{0,0.25,0.5}
\definecolor{redbrown}{rgb}{0.875,0.25,0.125}
\definecolor{darkgreen}{rgb}{0,0.5,0}

\renewcommand{\b}[1]{\ensuremath{\mathbf{#1}}}
\renewcommand{\H}{\ensuremath{\text{H}}}

\newcommand{\lr}{\ensuremath{\text{lr}}}
\newcommand{\sr}{\ensuremath{\text{sr}}}
\newcommand{\ee}{\ensuremath{\text{ee}}}

\newcommand{\HF}{\ensuremath{\text{HF}}}

\renewcommand{\d}{\ensuremath{\text{d}}}
\newcommand{\s}{\ensuremath{\text{s}}}

\newcommand{\x}{\ensuremath{\text{x}}}
\newcommand{\xc}{\ensuremath{\text{xc}}}

\DeclareMathOperator{\erf}{erf}

\renewcommand{\i}{\ensuremath{\text{i}}}

\begin{document}

\title{Linear-response time-dependent density-functional theory with local range separation: Core and valence resonances of the neon atom}

\author{Jari~van~Gog}
\affiliation{Laboratoire de Chimie Th\'eorique, Sorbonne Universit\'e and CNRS, F-75005 Paris, France}

\author{Felipe Zapata}
\affiliation{Departamento de Qu\'imica F\'isica, Universidad Complutense de Madrid, 28040 Madrid, Spain}

\author{Julien Toulouse}
\email{julien.toulouse@sorbonne-universite.fr}
\affiliation{Laboratoire de Chimie Th\'eorique, Sorbonne Universit\'e and CNRS, F-75005 Paris, France}

\date{July 17, 2026}

\begin{abstract}
We investigate range-separated hybrids (RSHs) and locally range-separated hybrids (LRSHs) for linear-response time-dependent density-functional theory (TDDFT) Sternheimer calculations of the photoionization spectrum of the Ne atom. This system constitutes a stringent test for approximate exchange-correlation treatments because it exhibits both valence and core resonances with very different energy scales. Building on previous work employing a simple one-parameter local range-separation function, we assess here a more flexible two-parameter range-separation function designed to improve the high-density limit. We compare photoionization spectra and resonance parameters obtained with RSHs and LRSHs. We find that the LRSH approach with the two-parameter range-separation function provides an overall satisfactory description of the photoionization spectrum, including energies for both valence and core resonances. The lifetimes of the 2s $\to n$p valence resonances are also reasonably reproduced since their decay does not involve double excitations. In contrast, the lifetimes of the 1s $\to n$p core resonances are overestimated by orders of magnitude because their Auger decay channels involve double excitations that are absent in adiabatic, single-determinant TDDFT. Obtaining more accurate resonance widths within linear-response range-separated TDDFT would require using multideterminant schemes and/or adding a frequency-dependent response kernel.
\end{abstract}

\maketitle

\section{Introduction}

Within density-functional theory (DFT), range-separated hybrids (RSHs) (see, e.g., Refs.~\onlinecite{IikTsuYanHir-JCP-01,YanTewHan-CPL-04,AngGerSavTou-PRA-05,GerAng-CPL-05a,VydScu-JCP-06}) constitute a widely used class of approximations for electronic-structure calculations. In their simplest variant, based on a range separation of the electron-electron interaction, they combine a long-range Hartree-Fock (HF) exchange contribution with a short-range exchange-correlation density functional. This class of approximations has proven particularly useful for linear-response time-dependent density-functional theory (TDDFT) calculations of excitation energies and response properties (see, e.g., Refs.~\onlinecite{TawTsuYanYanHir-JCP-04,PeaHelSalKeaLutTozHan-PCCP-06,LivBae-PCCP-07,SteKroBae-JCP-09,RebSavTou-MP-13,TouRebGouDobSeaAng-JCP-13}). 

Perhaps the most important limitation in standard RSHs is the use of a global range-separation parameter. To increase the flexibility of these approximations, locally range-separated hybrids (LRSHs), in which the range-separation parameter is replaced by a range-separation function of space, are presently actively developed~\cite{KruScuPerSav-JCP-08,HenJanScuSav-IJQC-09,AscKum-JCP-19,KlaBah-JCTC-20,MaiIkaNak-JCP-21,BruBahKum-JCP-22,BruBahKum-JCP-24,BruBahKum-JPCA-24,Mai-JCP-24}. By choosing the range-separation function as a functional of the density, LRSHs provide good accuracy for both core and valence properties~\cite{Mai-JCP-24}. In Refs.~\onlinecite{SchZapLevCanLupTou-JCP-22,TouSchZapLevCanLup-JCP-23}, a LRSH scheme with a simple range-separation function was implemented in a linear-response TDDFT Sternheimer framework and applied to the calculation of photoionization spectra of the beryllium and lithium atoms, including core resonances.

In the present work, we further explore the merits of the LRSH scheme for the calculation of the photoionization spectrum of the neon atom. The photoionization spectrum of this atom exhibits both core and valence resonances, with very different energy scales, and the description of these two types of resonances with similar accuracy requires the use of a position-dependent range-separation parameter. In comparison to our previous works~\onlinecite{SchZapLevCanLupTou-JCP-22,TouSchZapLevCanLup-JCP-23}, in which we used a simple one-parameter range-separation function introduced in Ref.~\onlinecite{KruScuPerSav-JCP-08}, we here test the use of the more flexible two-parameter range-separation function proposed in Ref.~\onlinecite{AscKum-JCP-19}, which helps to increase accuracy in high-density regions.

The paper is organized as follows. In Section~\ref{sec:theory}, we review the theory and the computational method. In Section~\ref{sec:results}, we provide and discuss the results. In particular, we explain how we choose the parameters of the range-separation functions based on the orbital energies, and we compare the photoionization spectra and resonance parameters obtained with the different methods. Finally, Section~\ref{sec:conclusion} contains our conclusions. Unless otherwise indicated, Hartree atomic units are assumed in this work.

\section{Theory and computational method}
\label{sec:theory}

\subsection{Ground-state locally range-separated hybrid scheme}
We consider an $N$-electron system with a closed-shell ground state. In the single-determinant LRSH scheme of Refs.~\onlinecite{SchZapLevCanLupTou-JCP-22,TouSchZapLevCanLup-JCP-23}, the ground-state orbitals $\{\psi_i\}$ and their associated energies $\{\varepsilon_i\}$ are found from the self-consistent Schr\"odinger-type equation
\begin{eqnarray}
\left( -\frac{1}{2} \nabla^2 + v_\text{ne}(\b{r}) + v_\H[\rho_0](\b{r}) + v_\xc^\sr[\rho_0](\b{r}) \right) \psi_i(\b{r})
\nonumber\\
+ \int v_\x^{\lr,\HF}[\gamma_0](\b{r},\b{r}') \psi_i(\b{r}') \d \b{r}' = \varepsilon_i \psi_i(\b{r}),
\label{RSH}
\end{eqnarray}
where $\gamma_0(\b{r},\b{r}') = 2\sum_{i=1}^{N/2} \psi_i(\b{r}) \psi_i^*(\b{r}')$ is the ground-state density matrix and $\rho_0(\b{r})=\gamma_0(\b{r},\b{r})$ is the ground-state density. In Eq.~\eqref{RSH}, $v_{\text{ne}}(\b{r})$ is the nuclei-electron potential, $v_\H[\rho_0](\b{r})=\int w_\ee(\b{r},\b{r}') \rho_0(\b{r}') \d \b{r}'$ is the Hartree potential, written with the Coulomb electron-electron interaction $w_\ee(\b{r},\b{r}')=1/|\b{r}-\b{r}'|$, $v_x^{\lr,\HF}[\gamma_0](\b{r},\b{r}')= (-1/2) w_\ee^\lr(\b{r},\b{r}') \gamma_0(\b{r},\b{r}')$ is the long-range (lr) nonlocal Hartree-Fock (HF) exchange potential, written with the long-range electron-electron interaction
\begin{eqnarray}
w_\ee^\lr(\b{r},\b{r}')=\frac{1}{2} \left[ \frac{\erf(\mu(\b{r})|\b{r}-\b{r}'|)}{|\b{r}-\b{r}'|} + \frac{\erf(\mu(\b{r}')|\b{r}-\b{r}'|)}{|\b{r}-\b{r}'|}\right],
\label{weelrerfmur}
\end{eqnarray}
where $\mu(\b{r})$ is the range-separation function, and $v_\xc^\sr[\rho_0](\b{r})$ is the complementary short-range (sr) exchange-correlation potential. For the latter term, we use the short-range local-density approximation (srLDA)
\begin{eqnarray}
v_\xc^{\sr}[\rho_0](\b{r}) = \left. \frac{\partial \bar{e}_{\xc,\text{UEG}}^\sr(\rho,\mu(\b{r}))}{\partial \rho} \right|_{\rho=\rho_0(\b{r})},
\label{vxcsr}
\end{eqnarray}
where $\bar{e}_{\xc,\text{UEG}}^\sr(\rho,\mu)$ is the complementary short-range exchange-correlation energy density of the uniform-electron gas (UEG) of density $\rho$, as parametrized in Ref.~\onlinecite{PazMorGorBac-PRB-06}. Note that in Eq.~\eqref{vxcsr} we use the srLDA functional constructed for a constant range-separation parameter $\mu$ but we evaluate it with the range-separation function $\mu(\b{r})$. 

In the special case where the range-separation function is chosen as a constant
\begin{eqnarray}
\mu_\text{RSH}(\b{r}) = \frac{\tilde{\mu}}{a_0},
\label{murRSH}
\end{eqnarray}
where $\tilde{\mu} \in [0,+\infty)$ is the adimensional range-separation parameter and $a_0=1$ a.u. is the Bohr radius, we recover the RSH scheme of Ref.~\onlinecite{AngGerSavTou-PRA-05}. We also consider two forms of range-separation functions expressed with the density $\rho(\b{r})$. The first one is
\begin{eqnarray}
\mu_\text{LRSH-K}(\b{r}) = \frac{\tilde{\mu}}{2} \frac{|\nabla \rho(\b{r})|}{\rho(\b{r})}.
\label{murK}
\end{eqnarray}
This form of range-separation function was used by Krukau (K) \textit{et al.}~\cite{KruScuPerSav-JCP-08} (see also Ref.~\onlinecite{TouColSav-JCP-05}) with a parameter of $\tilde{\mu} = 0.270$ optimized on atomization energies. It was also proposed in Ref.~\onlinecite{MaiIkaNak-JCP-21} to fix the parameter $\tilde{\mu}$ so as to recover the correct second-order density-gradient expansion of the exchange energy, leading to $\tilde{\mu} = 0.248$. Importantly, the range-separation function in Eq.~\eqref{murK} can be interpreted as an indicator of atomic electron shells~\cite{KohSavPre-JCP-91,NagMar-MP-97}. Assuming an exponentially decaying density in shell $i$, i.e. $\ln \rho(\b{r}) \approx -2\alpha_i |\b{r}|/a_0$ where $\alpha_i$ characterizes the radial decay of shell $i$, we have in this shell $\mu_\text{LRSH-K}(\b{r}) \approx \tilde{\mu} \, \alpha_i$. Hence, $\mu_\text{LRSH-K}(\b{r})$ is approximately a piecewise constant function, with $\mu_\text{LRSH-K}(\b{r}) \approx \tilde{\mu} \, Z$ in the 1s core region where $Z$ is the atomic number, and $\mu_\text{LRSH-K}(\b{r}) \approx \tilde{\mu} \, \sqrt{2 I}$ in the outer valence shell where $I$ is the ionization potential (IP). The second form of range-separation function that we consider is
\begin{eqnarray}
\mu_\text{LRSH-AK}(\b{r}) = \frac{\tilde{\mu}}{2} \frac{|\nabla \rho(\b{r})|}{\rho(\b{r})} \left[ 1 + \ln\left( 1+ c_{\text{hd}} \, a_0 \frac{|\nabla \rho(\b{r})|}{\rho(\b{r})} \right) \right],
\label{murAK}
\end{eqnarray}
where $c_{\text{hd}}$ is an adimensional parameter. This form of range-separation function was proposed in Ref.~\onlinecite{AscKum-JCP-19}, with the motivation that the logarithmic contribution in Eq.~\eqref{murAK} makes the total exchange-correlation energy tend to the full-range HF exchange energy in the high-density (hd) limit. Indeed, for nondegenerate Kohn-Sham ground states, exchange dominates over correlation when the density is scaled to the high-density limit, i.e. $\gamma^3 \rho(\gamma\b{r})$ and $\gamma \to \infty$,~\cite{Lev-PRA-91} and it is thus preferable in this limit to use exact or HF exchange rather than an approximate density functional. In Ref.~\onlinecite{BruBahKum-JPCA-24}, it was found that optimization on atomization energies and reaction barrier heights gives the parameters $\tilde{\mu}=0.230$ and $c_{\text{hd}}=0.0232$, whereas optimization on $\Delta$SCF IPs gives the parameters $\tilde{\mu}=0.440$ and $c_{\text{hd}}=0.0495$. In practice, the logarithmic contribution in Eq.~\eqref{murAK} is mostly significant in the high-density regions, so that it increases the value of the range-separation function taken in core regions while leaving it almost unchanged in valence regions. We will refer to the specific LRSH schemes obtained by using the range-separation functions in Eqs.~\eqref{murK} and~\eqref{murAK} as LRSH-K and LRSH-AK, respectively.

Even though one can use the self-consistent LRSH ground-state density in the range-separation functions~\cite{KlaBah-JCTC-20,MaiIkaNak-JCP-21,BruBahKum-JCP-22,BruBahKum-JCP-24,BruBahKum-JPCA-24,Mai-JCP-24}, in the present work for simplicity we simply use the fixed HF ground-state density in Eqs.~\eqref{murK} and~\eqref{murAK}. Note that the HF ground-state density is used only for defining the range-separation function $\mu(\b{r})$, but the LRSH orbitals and density used in the Hartree, exchange and correlation terms are properly calculated self-consistently. For $\tilde{\mu}=0$, the present RSH and LRSH schemes reduce to standard Kohn-Sham LDA. For $\tilde{\mu}\to\infty$, they reduce to standard HF.

\subsection{Linear-response Sternheimer scheme for photoionization spectra}
After the ground-state calculation, the first-order orbital responses $\psi_i^{(+)}(\b{r},\omega)$ and $\psi_i^{(-)}(\b{r},\omega)$ to a dipole interaction with a monochromatic electric field of frequency $\omega$ are calculated from the linear-response adiabatic Sternheimer equations (see, e.g., Refs.~\onlinecite{MahSub-BOOK-90,SteDecLis-JPB-95,SteDec-JCP-00,YabNakIwaBer-PSS-06,AndBotMarRub-JCP-07,StrLehRubMarLou-INC-12,HofSchKum-JCP-18,HofSchKum-PRA-19,HofKum-JCP-20}) adapted to range-separated hybrids~\cite{SchZapLevCanLupTou-JCP-22,TouSchZapLevCanLup-JCP-23}
\begin{align}
\left(\varepsilon_i \pm \omega \right) \psi_i^{(\pm)}(\b{r},\omega) = \phantom{xxxxxxxxxxxxxxxxxxxxx}
\nonumber\\
\left( -\frac{1}{2} \nabla^2 + v_\text{ne}(\b{r}) + v_\H[\rho_0](\b{r}) + v_\xc^\sr[\rho_0](\b{r}) \right) \psi_i^{(\pm)}(\b{r},\omega)
\nonumber\\
+ \int v_\x^{\lr,\HF}[\gamma_0](\b{r},\b{r}') \psi_i^{(\pm)}(\b{r}',\omega) \d \b{r}'  \phantom{xxxxxxxxxxxx}
\nonumber\\
+ \left( v_\H[\rho^{(\pm)}](\b{r}) + v_\xc^{\sr,(1)}[\rho^{(\pm)}](\b{r}) \right) \psi_i(\b{r}) \phantom{xxxxxxxxx}
\nonumber\\
+ \int v_\x^{\lr,\HF}[\gamma^{(\pm)}](\b{r},\b{r}') \psi_i(\b{r}') \d \b{r}' 
+v_\text{ext}^{(1)}(\b{r}) \psi_i(\b{r}),
\label{Sternheimerrealspace}
\end{align}
where we have introduced the first-order density matrix
\begin{eqnarray}
\gamma^{(\pm)}(\b{r},\b{r}',\omega) = 2\sum_{i=1}^{N/2} \left[\psi_i^{(\pm)}(\b{r},\omega) \psi_i^*(\b{r}') + \psi_i(\b{r}) \psi_i^{(\mp)*}(\b{r}',\omega) \right],
\nonumber\\
\label{gammapm}
\end{eqnarray}
and the corresponding first-order density $\rho^{(\pm)}(\b{r},\omega) = \gamma^{(\pm)}(\b{r},\b{r},\omega)$. In Eq.~\eqref{Sternheimerrealspace}, the first-order potentials are $v_\H[\rho^{(\pm)}](\b{r})=\int w_\ee(\b{r},\b{r}') \rho^{(\pm)}(\b{r}',\omega) \d \b{r}'$, $v_\xc^{\sr,(1)}[\rho^{(\pm)}] = \int f_\xc^{\sr}[\rho_0](\b{r},\b{r}') \rho^{(\pm)}(\b{r}',\omega) \d \b{r}'$ where $f_\xc^{\sr}[\rho_0](\b{r},\b{r}') = \delta v_\xc^\sr[\rho_0](\b{r})/\delta \rho(\b{r'})$ is the adiabatic short-range exchange-correlation response kernel, $v_\x^{\lr,\HF}[\gamma^{(\pm)}](\b{r},\b{r}') = (-1/2) w_\ee^\lr(\b{r},\b{r}') \gamma^{(\pm)}(\b{r},\b{r}')$, and $v_\text{ext}^{(1)}(\b{r}) = \b{r} \cdot \b{e}$  where $\b{e}$ is the unit polarization vector of the electric field. For the LRSH scheme, note that Eq.~\eqref{Sternheimerrealspace} corresponds to the linear response of the orbitals and density used in the Hartree, exchange and correlation terms but for a fixed range-separation function $\mu(\b{r})$.

The photoionization cross section is then calculated as
\begin{eqnarray}
\sigma(\omega) = \frac{4\pi \omega}{c} \text{Im}[\alpha(\omega)],
\label{sigma}
\end{eqnarray}
where $c = 137.036$ a.u. is the speed of light and $\alpha(\omega)$ is the spherically averaged dipole polarizability given by
\begin{eqnarray}
\alpha(\omega) = - \frac{1}{3} \sum_{a\in \{x,y,z\}} \int (\b{r} \cdot \b{u}_a) \; \rho^{(+)} (\b{r},\omega) \d\b{r},
\label{alpha}
\end{eqnarray}
where $\b{u}_a$ is the unit vector along the direction $a$.

\begin{figure}
\includegraphics[scale=0.35,angle=0]{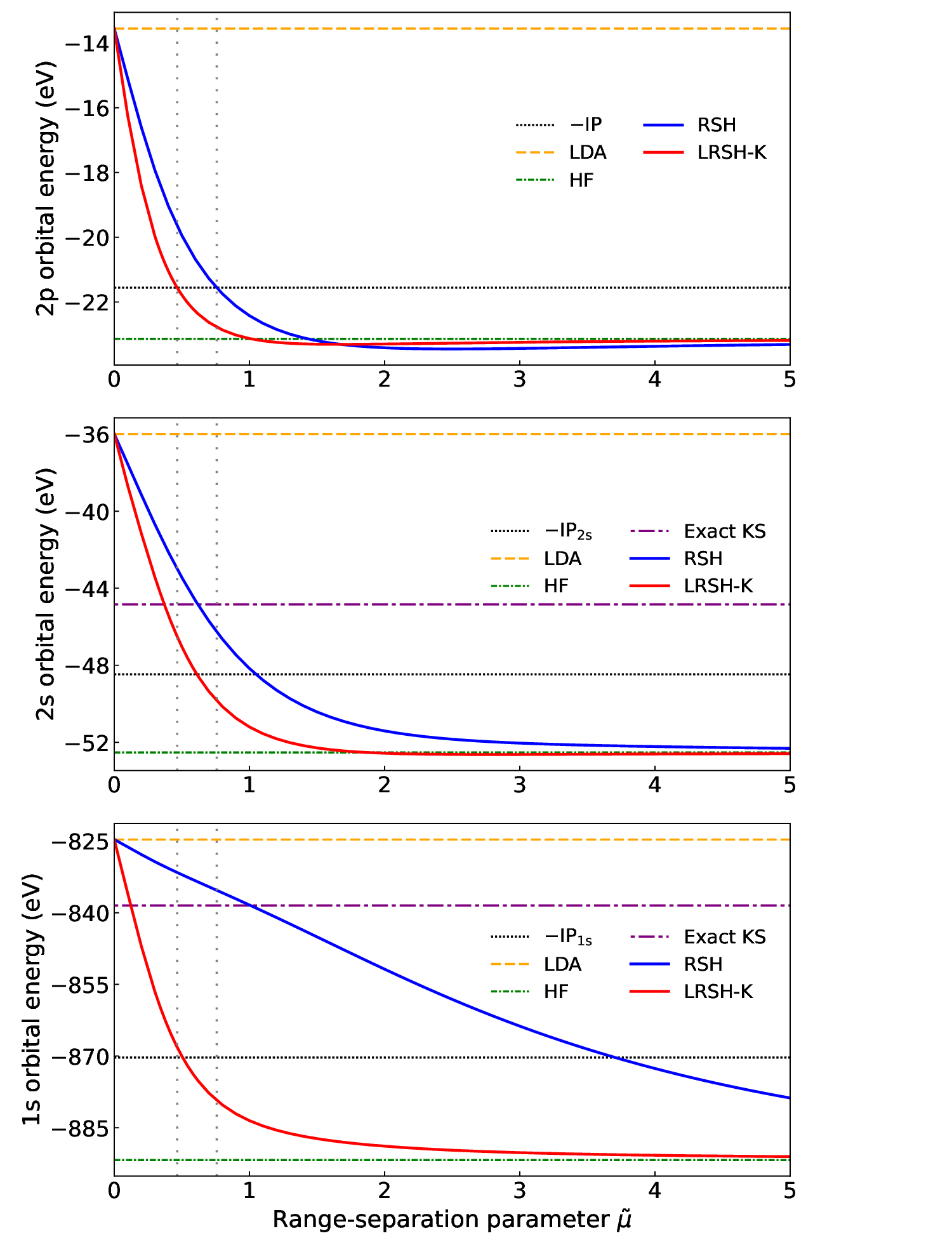}
	\caption{RSH and LRSH-K ($c_{\text{hd}}=0$) 1s, 2s and 2p orbital energies of the Ne atom as a function of the adimensional range-separation parameter $\tilde{\mu}$. The opposite of the experimental IP (21.56 eV)~\cite{NIST-INC-24} and of the experimental 2s and 1s ionization edges (IP$_\text{2s}=48.47$ and IP$_\text{1s}=870.31$ eV)~\cite{JolBomEye-ADNDT-84} are indicated as references. The exact KS 2s and 1s orbital energies (44.84 and 838.48 eV)~\cite{ChoGriBae-JCP-02} are also shown, as well as the orbital energies in the two limit cases: LDA ($\tilde{\mu}=0$) and HF ($\tilde{\mu}\rightarrow \infty$). The vertical dotted lines are situated at $\tilde{\mu}=0.466$ and $\tilde{\mu}=0.758$, corresponding to the optimal value of the range-separation parameter such that the 2p orbital energy matches the opposite of the experimental IP for the LRSH-K and RSH schemes, respectively.}
\label{fig:orbitalenergies}
\end{figure}

\begin{figure}
\includegraphics[scale=0.35,angle=0]{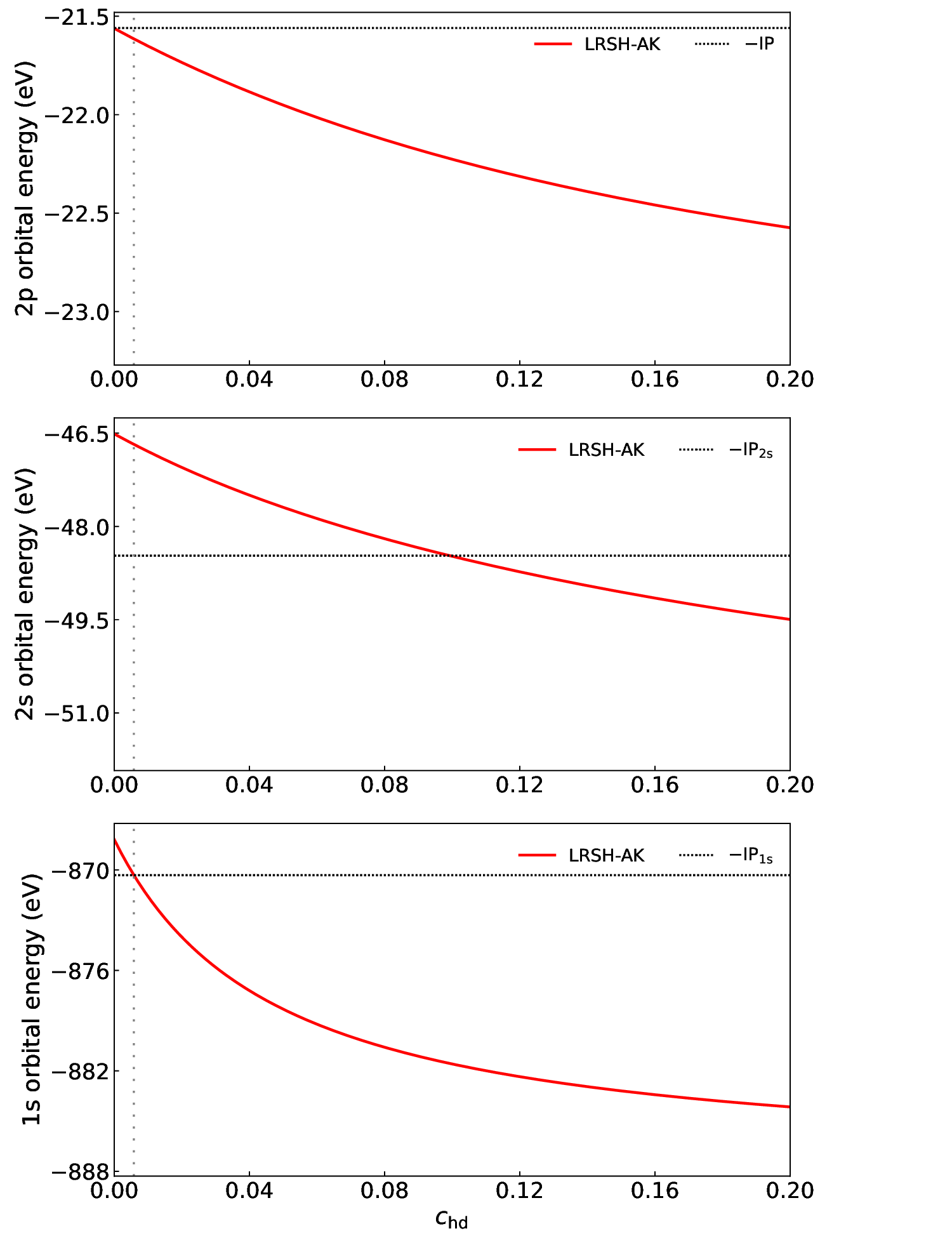}
	\caption{LRSH-AK 1s, 2s and 2p orbital energies of the Ne atom as a function of parameter $c_{\text{hd}}$ in Eq.~\eqref{murAK} at fixed range-separation parameter $\tilde{\mu}=0.466$. The opposite of the experimental IP (21.56 eV)~\cite{NIST-INC-24} and of the experimental 2s and 1s ionization edges (IP$_\text{2s}=48.47$ and IP$_\text{1s}=870.31$ eV)~\cite{JolBomEye-ADNDT-84} are indicated indicated as references. The vertical dotted line is situated at $c_{\text{hd}}=0.0058$, corresponding to the optimal value of $c_{\text{hd}}$ such that the 1s orbital energy matches the opposite of the experimental 1s ionization edge.}
\label{fig:orbitalenergiescHD}
\end{figure}

\begin{figure}
\includegraphics[scale=0.5,angle=0]{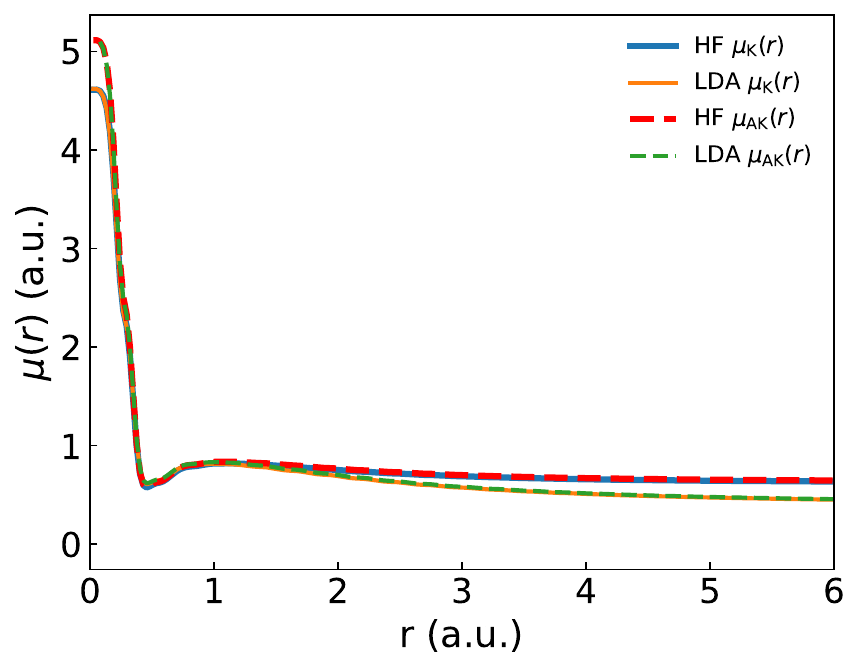}
\caption{LRSH-K and LRSH-AK range-separation functions $\mu_\text{LRSH-K}(\b{r})$ [Eq.~\eqref{murK}] and $\mu_\text{LRSH-AK}(\b{r})$ [Eq.~\eqref{murAK}] of the Ne atom as a function of the electron-nucleus distance $r$, calculated with fixed ground-state HF and LDA densities. The parameters are $\tilde{\mu}=0.466$ for LRSH-K, and $\tilde{\mu}=0.466$ and $c_{\text{hd}}=0.0058$ for LRSH-AK.}
\label{fig:range_separation_functions_comparison}
\end{figure}

\begin{figure*}
\includegraphics[scale=0.4,angle=0]{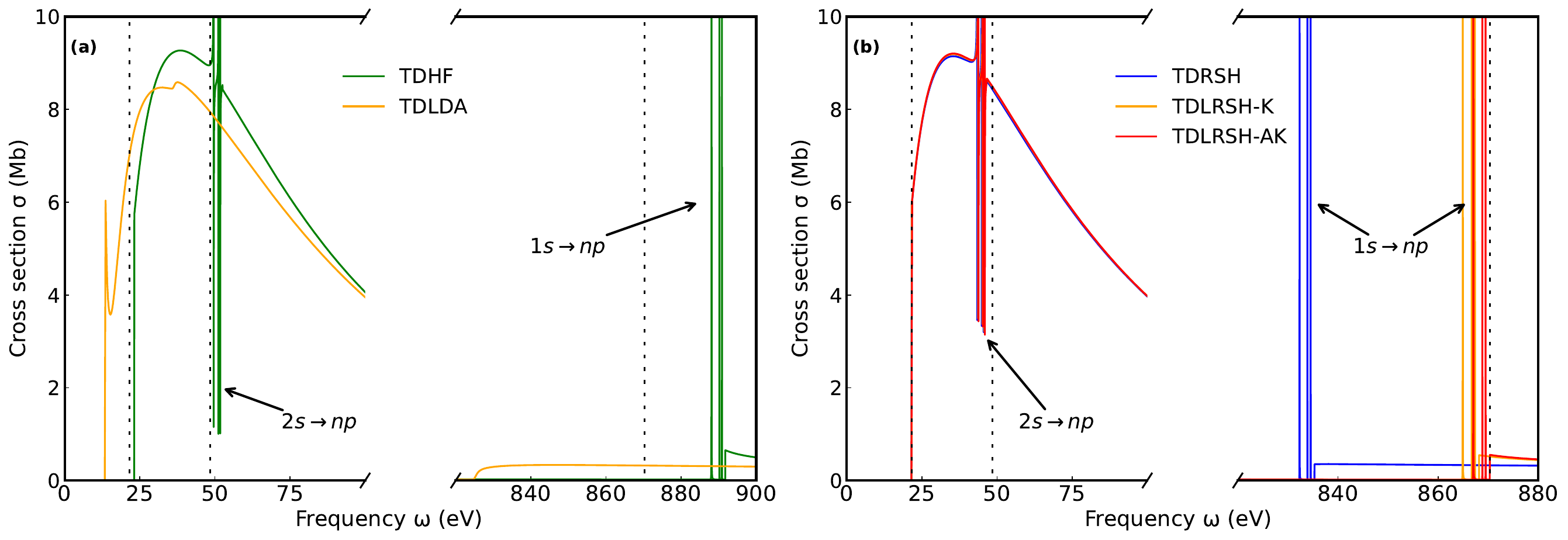}
\caption{Photoionization cross section of the Ne atom calculated by (a) TDHF and TDLDA, and by (b) TDRSH ($\tilde{\mu}=0.758$), TDLRSH-K ($\tilde{\mu}=0.466$), and TDLRSH-AK ($\tilde{\mu}=0.466$ and $c_{\text{hd}}=0.0058$). The vertical dashed lines correspond to the experimental IP (21.56 eV)~\cite{NIST-INC-24}, the 2s ionization edge (48.47 eV), and the 1s ionization edge (870.31 eV)~\cite{JolBomEye-ADNDT-84}.}
\label{fig:photoionization}
\end{figure*}

\subsection{Computational details}
For atoms, we exploit spherical symmetry and we write the occupied orbitals as
\begin{eqnarray}
\psi_{i}(\b{r}) = \frac{R_i(r)}{r} Y_{\ell_i}^{m_i}(\theta,\phi),
\end{eqnarray}
with radial functions $R_i$ and spherical harmonics $Y_{\ell_i}^{m_i}$. Choosing a $z$-polarized electric field, i.e. $v_\text{ext}^{(1)}=  z $, 
the corresponding first-order orbital changes are of the form
\begin{eqnarray}
\psi_{i}^{(\pm)}(\b{r},\omega) = \sum_{\ell\in {\cal L}_{i}} \frac{R_{i,\ell}^{(\pm)}(r,\omega)}{r} Y_{\ell}^{m_i}(\theta,\phi),
\label{psijrmratoms}
\end{eqnarray}
with ${\cal L}_i  = \{ \ell_i -1, \ell_i +1 \}$ for $\ell_i\ge 1$ and ${\cal L}_i  = \{ \ell_i +1 \}$ for $\ell_i=0$. The radial functions $R_i$ and $R_{i,\ell}^{(\pm)}$ are expanded in a basis set of $M_\s$ B-spline functions~\cite{Boo-BOOK-78,BacCorDecHanMar-RPP-01} of order $k_\s$ and maximal radius of $r_{\text{max}}$, corresponding to a linear grid of spacing $\Delta r = r_{\text{max}}/(M_\s-k_\s+1)$ bohr. At $r=r_{\text{max}}$, we impose Neumann boundary conditions for $R_{i}(r)$
\begin{align}
\left.  \frac{\d R_{i}(r)}{\d r} \right|_{r=r_{\text{max}}} = 0,
\end{align}
and either Neumann or outgoing-spherical-wave Robin boundary conditions for $R_{i,\ell}^{(\pm)}(r,\omega)$ depending whether it corresponds to a bound or a continuum state~\cite{SchZapLevCanLupTou-JCP-22}
\begin{align}
\left.  \frac{\d R_{i,\ell}^{(\pm)}(r,\omega)}{\d r} \right|_{r=r_{\text{max}}} = b_{i,\ell}(\pm\omega) R_{i,\ell}^{(\pm)}(r_{\text{max}}),
\end{align}
where
\begin{eqnarray}
b_{i,\ell}(\omega) = 
\begin{cases}
\left. \dfrac{\i \; \d f_{i,\ell}(r,\omega)/\d r + \d g_{i,\ell}(r,\omega)/\d r}{\i f_{i,\ell}(r,\omega) + g_{i,\ell}(r,\omega)} \right|_{r=r_\text{max}} &
\text{if $\omega \geq-\varepsilon_i$}\\
0 & \text{if $\omega < -\varepsilon_i$}
\end{cases}
\nonumber\\
\label{blargeomega}
\end{eqnarray}
with $f_{i,\ell}(r,\omega) = F_{\ell}(- Z_\text{eff}/k_i(\omega),k_i(\omega) r)$ and $g_{i,\ell}(r,\omega) = G_{\ell}(-Z_\text{eff}/k_i(\omega),k_i(\omega) r)$, where $F_{\ell}$ and $G_{\ell}$ are the regular and irregular Coulomb functions~\cite{AbrSte-BOOK-83}, $k_i(\omega)=\sqrt{2(\varepsilon_i + \omega)}$ is the free-electron momentum, and $Z_\text{eff}=Z-N+\zeta$ is the effective charge seen by the free electron, with the nuclear charge $Z$ and the HF exchange contribution $\zeta$ (we have $\zeta=1$ in the presence of long-range HF exchange and $\zeta =0$ in the absence of HF exchange). In practice, we solve the linear-response Sternheimer equations written in the B-spline basis set~\cite{SchZapLevCanLupTou-JCP-22,TouSchZapLevCanLup-JCP-23}.

We apply the present approach to the neon atom ($N=10)$ with a homemade program, using the B-spline parameters $M_\s=100$, $k_\s=8$, and $r_{\text{max}} = 25$ bohr, corresponding to a grid spacing of $\Delta r = 0.269$ bohr. We have checked that the total and orbital energies for the HF and LDA methods are converged with respect to $\Delta r$ and $r_{\text{max}}$. We have also checked that the photoionization spectra for the TDHF and TDLDA methods are well converged with respect to these parameters. We calculate the short-range exchange-correlation potential $v_\xc^\sr$ and kernel $f_\xc^\sr$ from the short-range exchange-correlation LDA functional of Ref.~\onlinecite{PazMorGorBac-PRB-06} using the same range-separation function $\mu(\b{r})$ used in the long-range HF exchange potential. The time-dependent linear-response methods based on the RSH, LRSH-K and LRSH-AK schemes will be referred to as TDRSH, TDLRSH and TDLRSH-AK, respectively. For $\tilde{\mu}=0$ the TDRSH, TDLRSH-K and TDLRSH-AK methods reduce to standard linear-response time-dependent local-density approximation (TDLDA), while for $\tilde{\mu} \to \infty$ they reduce to standard linear-response time-dependent Hartree--Fock (TDHF).

\section{Results and discussion}
\label{sec:results}

Figure \ref{fig:orbitalenergies} shows the RSH and LRSH-K 1s, 2s and 2p occupied orbital energies as a function of the range-separation parameter $\tilde{\mu}$. As references, we report the experimental ionization potential (IP), 21.56 eV~\cite{NIST-INC-24}, as well as the 2s and 1s experimental IPs (defined as the 2s and 1s ionization edges), IP$_\text{2s}=48.47$ and IP$_\text{1s}=870.31$ eV~\cite{JolBomEye-ADNDT-84}, respectively. The exact KS 2p orbital energy is identical to the opposite of the IP. By contrast, the exact KS 2s and 1s orbital energies (44.84 and 838.48 eV)~\cite{ChoGriBae-JCP-02} are not identical to $-\text{IP}_\text{2s}$  and  $-\text{IP}_\text{1s}$. In exact linear-response TDDFT, the exact 2s and 1s IPs are nevertheless recovered, but this requires the exact exchange-correlation response kernel to exhibit a singular behavior. For the RSH and LRSH methods considered in the present work, the approximate exchange-correlation response kernels cannot change the IPs (see, e.g., Refs.~\onlinecite{BarLotSch-JCP-05,SchZapLevCanLupTou-JCP-22}), and thus to obtain the correct IPs in TDRSH and TDLRSH, each occupied orbital energy must be equal to the opposite of the corresponding IP. This justifies the comparison of each occupied orbital energy with the opposite of the corresponding experimental IP. If TDRSH and TDLRSH give accurate IPs, then we can also expect accurate resonance energies to resonant states just below the different ionization edges. 

Within the RSH and LRSH schemes, the value of the range-separation parameter $\tilde{\mu}$, which governs the decomposition of the electron-electron interaction, must be chosen prior to any calculation. In the spirit of the so-called optimally tuned range-separated hybrids~\cite{LivBae-PCCP-07,SteKroBae-JACS-09,SteKroBae-JCP-09}, we define the optimal parameter $\tilde{\mu}$ such that the 2p orbital energy equals the opposite of the experimental IP. This results in an optimal value of $\tilde{\mu}=0.758$ for RSH and $\tilde{\mu}=0.466$ for LRSH-K. These values will be used in the calculations of the RSH and LRSH-K photoionization spectra. The crucial advantage of LRSH-K over RSH is that that the value $\tilde{\mu}=0.466$ is also quite close to being optimal to align the LRSH-K 2s and 1s orbital energies with the opposite of the 2s and 1s experimental IPs (the optimal values of $\tilde{\mu}$ for the LRSH-K 2s and 1s orbital energies are $0.611$ and $0.506$). By contrast, the value $\tilde{\mu}=0.758$ is far from being optimal for the RSH 2s and 1s orbital energies (the optimal values of $\tilde{\mu}$ for the RSH 2s and 1s orbital energies are $1.048$, and $3.705$). Thus, thanks to the use of the local range-separation function $\mu_\text{LRSH-K}(\b{r})$ in Eq.~\eqref{murK}, LRSH-K is able to treat all the core and valence occupied orbitals on an approximately equal footing with a single parameter $\tilde{\mu}$.

\begin{figure*}[t]
\includegraphics[scale=0.4,angle=0]{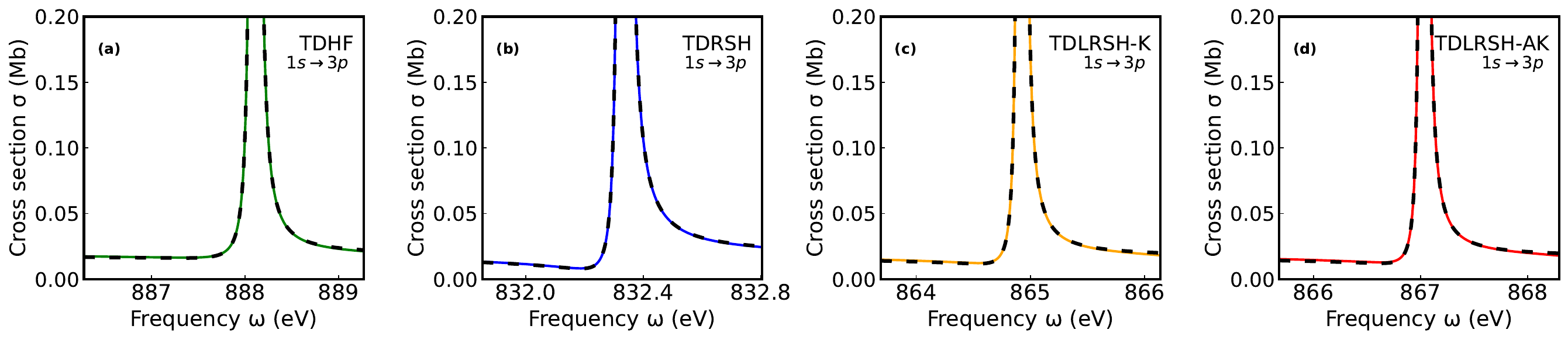}
\caption{Core resonance 1s $\to$ 3p of the Ne atom calculated by (a) TDHF, (b) TDRSH ($\tilde{\mu}=0.758$), (c) TDLRSH-K ($\tilde{\mu}=0.466$), and (d) TDLRSH-AK ($\tilde{\mu}=0.466$, $c_{\text{hd}}=0.0058$). The dashed lines are fits using Eq.~(\ref{sigmafit}) with the parameters given in Table~\ref{tab:resonance}.}
\label{fig:resonance_core1}
\end{figure*}

\begin{figure*}
\includegraphics[scale=0.4,angle=0]{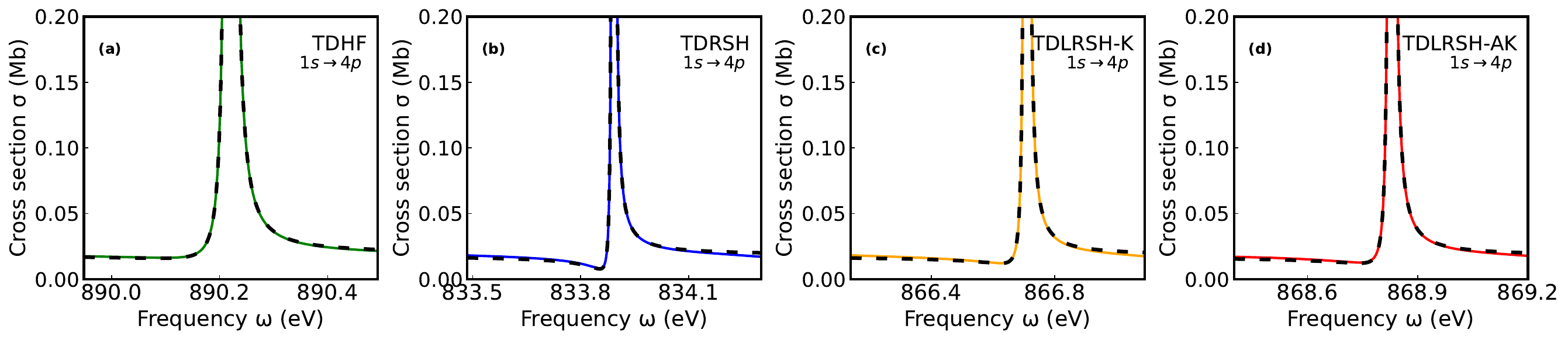}
\caption{Core resonance 1s $\to$ 4p of the Ne atom calculated by (a) TDHF, (b) TDRSH ($\tilde{\mu}=0.758$), (c) TDLRSH-K ($\tilde{\mu}=0.466$), and (d) TDLRSH-AK ($\tilde{\mu}=0.466$, $c_{\text{hd}}=0.0058$). The dashed lines are fits using Eq.~(\ref{sigmafit}) with the parameters given in Table~\ref{tab:resonance}.}
\label{fig:resonance_core2}
\end{figure*}

\begin{figure*}
\includegraphics[scale=0.4,angle=0]{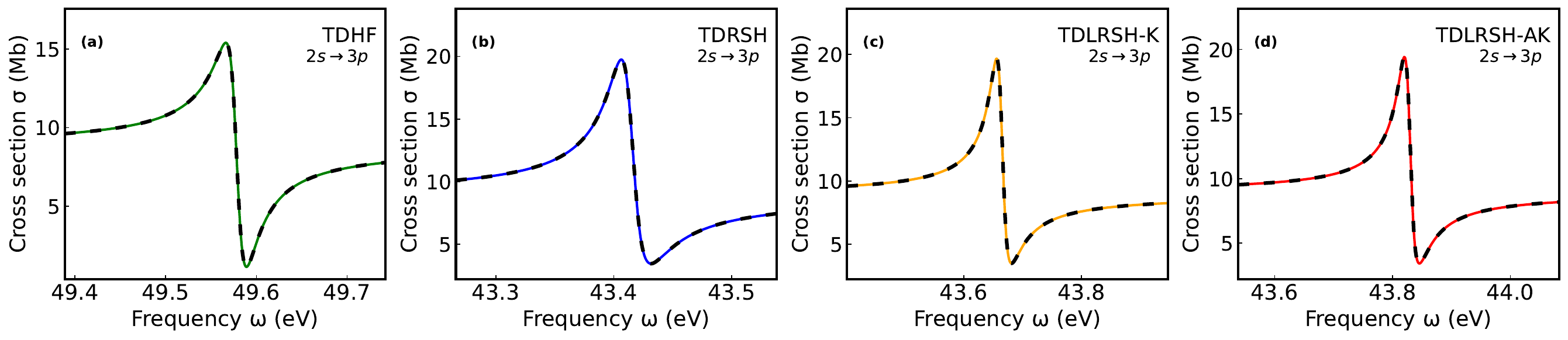}
\caption{Valence resonance 2s $\to$ 3p of the Ne atom calculated by (a) TDHF, (b) TDRSH ($\tilde{\mu}=0.758$), TDLRSH-K ($\tilde{\mu}=0.466$), and (d) TDLRSH-AK ($\tilde{\mu}=0.466$, $c_{\text{hd}}=0.0058$). The dashed lines are fits using Eq.~(\ref{sigmafit}) with the parameters given in Table~\ref{tab:resonance}.}
\label{fig:resonance_valence1}
\end{figure*}

\begin{figure*}
\includegraphics[scale=0.4,angle=0]{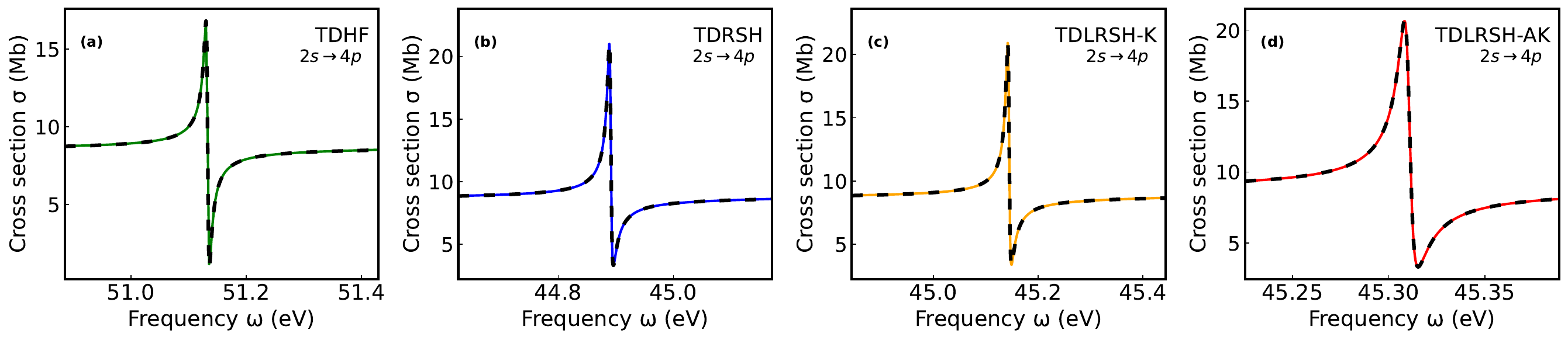}
\caption{Valence resonance 2s $\to$ 4p of the Ne atom calculated by (a) TDHF, (b) TDRSH ($\tilde{\mu}=0.758$), (c) TDLRSH-K ($\tilde{\mu}=0.466$), and (d) TDLRSH-AK ($\tilde{\mu}=0.466$, $c_{\text{hd}}=0.0058$). The dashed lines are fits using Eq.~(\ref{sigmafit}) with the parameters given in Table~\ref{tab:resonance}.}
\label{fig:resonance_valence2}
\end{figure*}

Using the fixed optimal value of the range-separation parameter $\tilde{\mu}=0.466$ determined for LRSH-K, we now explore the effect of the parameter $c_{\text{hd}}$ in the range-separation function $\mu_\text{LRSH-AK}(\b{r})$ in Eq.~\eqref{murAK} for LRSH-AK. As seen in Figure~\ref{fig:orbitalenergiescHD}, starting from $c_{\text{hd}} = 0$ and increasing $c_{\text{hd}}$ leads to a significant change of the 1s orbital energy, while the 2s and 2p orbital energies are less influenced by this parameter, which was expected since the logarithmic term in Eq.~\eqref{murAK} is more significant in high-density regions. Accordingly, we choose to determine $c_{\text{hd}}$ by matching the 1s orbital energy to the opposite of the 1s experimental IP, giving $c_{\text{hd}}=0.0058$. With this value of $c_{\text{hd}}$, the 2p orbital energy remains very close to the opposite of the exact IP, and the 2s orbital energy is slightly closer to the opposite of the 2s experimental IP. This value of $c_{\text{hd}}$ will be used in the calculation of the LRSH-AK photoionization spectrum.
Comparison with previous works~\cite{SchZapLevCanLupTou-JCP-22,TouSchZapLevCanLup-JCP-23,BruBahKum-JPCA-24} suggests that the optimal value for the parameter $\tilde{\mu}$ for LRSH when targeting IPs is well transferable from systems to systems. However, the transferability of the optimal value of the parameter $c_{\text{hd}}$ is an open question.

As mentioned before, the fixed ground-state HF density was used in the range-separation functions in Eqs.~\eqref{murK} and~\eqref{murAK}. To estimate the dependence of the range-separation functions on the density used, we compare in Figure~\ref{fig:range_separation_functions_comparison} the range-separation functions $\mu_\text{LRSH-K}(\b{r})$ and $\mu_\text{LRSH-AK}(\b{r})$ calculated with HF and LDA ground-state densities. As can be seen from this figure, at short electron-nucleus distances the range-separation functions are not influenced by the choice of the density. Moreover, at these distances the high-density-limit correction has a significant impact, increasing the value of the range-separation function by almost 0.5 a.u.. At large electron-nucleus distances the range-separation functions give roughly a plateau with approximately the value $\tilde{\mu} \sqrt{2I}$. This plateau is slightly higher when using the HF density, which is expected since the ionization potential $I$ is larger in HF than in LDA. We expect that the range-separation functions calculated from the self-consistent LRSH density would be close to the ones calculated with the HF density.

The photoionization spectra for the TDHF, TDLDA, TDRSH, TDLRSH-K, and TDLRSH-AK methods are shown in Figure~\ref{fig:photoionization}. The TDHF photoionization spectrum starts at a slightly too large ionization threshold (23.14 eV), with a non-zero cross section (about 5.7 Mb) at the threshold. It also significantly overestimates the positions of the 2s and 1s ionization edges (giving 52.52 and 891.66 eV, respectively). TDHF correctly gives the two resonance series $2\text{s}\rightarrow n\text{p}$ and $1\text{s}\rightarrow n\text{p}$ ($n \geq 3)$, converging to the 2s and 1s ionization edges, respectively. The TDLDA photoionization spectrum starts at a much too small ionization threshold (13.55 eV), with a zero cross section, as expected from the Wigner-threshold law~\cite{Wig-PR-48,SadBohCavEsrFabMacRau-JPB-00} for potentials lacking a long-range Coulomb tail $-1/r$. The TDLDA spectrum shows a sharp peak just above the ionization threshold in the energy range that should correspond to the  $2\text{p} \rightarrow 4\text{s}$ transition, which is incorrectly in the continuum. TDLDA also severely underestimates the positions of the 2s and 1s ionization edges (35.99 and 824.55 eV), which can be attributed to the LDA self-interaction error in the description of the 2s and 1s orbitals. Furthermore, TDLDA does not support any virtual bound p orbitals, and consequently the $2\text{s}\rightarrow n\text{p}$ and $1\text{s}\rightarrow n\text{p}$ resonances are completely absent in the spectrum. 

Due to our choice of the range-separation parameter, the TDRSH, TDLRSH-K, and TDLRSH-AK photoionization spectra start at the exact or nearly exact value of the ionization threshold (21.56, 21.56, and 21.61 eV, respectively). The TDRSH spectrum shows 2s and 1s ionization edges at 46.26 and 835.29 eV, the latter value being a severe underestimation. The TDLRSH-K spectrum shows 2s and 1s ionization edges at 46.51 and 868.18 eV. The TDLRSH-AK spectra gives 2s and 1s ionization edges at 46.68 and 870.30 eV, the latter value being nearly the exact value due to our choice of the high-density parameter $c_\text{hd}$. TDRSH, TDLRSH-K, and TDLRSH-AK all give resonance series $2\text{s}\rightarrow n\text{p}$ and $1\text{s}\rightarrow n\text{p}$.

The first two resonances of each series calculated by TDHF, TDRSH, TDLRSH-K, and TDLRSH-AK are plotted in Figures~\ref{fig:resonance_core1} to~\ref{fig:resonance_valence2}. The cross section follows a characteristic asymmetric Fano lineshape which can be fitted to the analytical expression~\cite{FanCoo-PR-65,SteDecLis-JPB-95}
\begin{equation}
\sigma(\omega) = \sigma_0 (1+a \, \epsilon(\omega)) \left[ \rho^2 \frac{(q+\epsilon(\omega))^2}{1+\epsilon(\omega)^2} -\rho^2+1\right],
\label{sigmafit}
\end{equation}
with $\epsilon(\omega) = 2(\omega-E_\text{R})/\Gamma$. In this expression, $E_\text{R}$ is the resonance energy, $\Gamma$ is the resonance width (or inverse lifetime), $q$ is the asymmetry Fano parameter, $\sigma_0$ is the total background cross section, $a$ is a coefficient for the total background linear drift, and $\rho^2$ is the ratio between the background cross section for transitions to continuum states that interact with the discrete resonant state and the total background cross section. The fitted parameters obtained from TDHF, TDRSH, TDLRSH-K, and TDLRSH-AK are given in Table~\ref{tab:resonance}, together with some reference parameters obtained from experiments~\cite{MulBerBorBuhHelHolKilKluMarRicSelVieChi-AJ-17,CodMadEde-PR-67,SchDomPutGutKaiMieGre-PRA-96} or from accurate calculations~\cite{SchDomPutGutKaiMieGre-PRA-96,MarKliKjeLinGonArgMar-PRA-17,SkoKry-JCP-21b,AlnHamAbu-CPI-25}.

For the two core resonances, TDHF, TDRSH, and TDLRSH give quite different resonance energies $E_\text{R}$, TDLRSH-AK giving the most accurate energies, 867.0 eV and 868.8 eV, compared to the experimental values, 867.3 and 868.9 eV, for the 1s $\to$ 3p and 1s $\to$ 4p resonances, respectively. This was expected since the parameter $c_\text{hd}$ in TDLRSH-AK was optimized on the 1s ionization edge. TDHF, TDRSH, and TDLRSH all dramatically underestimate the core resonance widths $\Gamma$ by two orders of magnitude or more. It is known that, in the Ne atom, the 1s $\to$ 3p resonance decays predominantly by the spectator Auger process~\cite{HayMurMorShiYagKoi-JPB-95,KivHeiJurAliNomAksAks-JESRP-01,SkoKry-JCP-21b,TenVosBokDecCor-JCTC-22}
\begin{equation}
1\text{s}2\text{s}^2 2\text{p}^6 3\text{p} \to 1\text{s}^2(2\text{s}2\text{p})^6 3\text{p} + \text{e}^-,
\end{equation}
while the 1s $\to$ 4p resonance decays predominantly by shake-up Auger processes~\cite{HayMurMorShiYagKoi-JPB-95,KivHeiJurAliNomAksAks-JESRP-01}
\begin{equation}
1\text{s}2\text{s}^2 2\text{p}^6 4\text{p} \to 1\text{s}^2(2\text{s}2\text{p})^6 5\text{p} + \text{e}^-.
\end{equation}
The states $1\text{s}^2(2\text{s}2\text{p})^6 3\text{p}$ and $1\text{s}^2(2\text{s}2\text{p})^6 5\text{p}$ are double excitations with respect to the ground-state configuration $1\text{s}^2 2\text{s}^2 2\text{p}^6$ and are not explicitly taken into account in linear-response TDHF or adiabatic single-determinant TDDFT. This explains why TDHF, TDRSH, and TDLRSH cannot properly describe the lifetimes of these resonances.

\begin{table*}
\caption{Resonance energy $E_\text{R}$, resonance width $\Gamma$, Fano asymmetric parameter $q$, total background cross section $\sigma_0$, background ratio parameter $\rho^2$, background linear drift $a$, and maximum value of the cross section at the resonance energy $\sigma(E_\text{R})$ for the 1s$\to$3p and 1s$\to$4p core resonances and the 2s$\to$3p and 2s$\to$4p valence resonances of the Ne atom calculated by TDHF, TDRSH ($\tilde{\mu}=0.758$), TDLRSH-K ($\tilde{\mu}=0.466$), and TDLRSH-AK ($\tilde{\mu}=0.466$, $c_{\text{hd}}=0.0058$). As references, we also report experimental values~\cite{MulBerBorBuhHelHolKilKluMarRicSelVieChi-AJ-17,CodMadEde-PR-67,SchDomPutGutKaiMieGre-PRA-96} and accurate results from equation-of-motion coupled-cluster singles doubles (EOM-CCSD)~\cite{SkoKry-JCP-21b}, relativistic configuration interaction (RCI)~\cite{AlnHamAbu-CPI-25}, R-matrix~\cite{SchDomPutGutKaiMieGre-PRA-96}, and extended configuration interaction (XCI)~\cite{MarKliKjeLinGonArgMar-PRA-17} calculations. The numbers in parenthesis are estimated uncertainties.}

\label{tab:resonance}
\begin{tabular}{l c c c c c c c}
\hline\hline
                       & $E_\text{R}$ (eV) & $\Gamma$ (meV) & $q$ & $\sigma_0$  (Mb)  & $\rho^2$ & $a$  & $\sigma(E_\text{R})$ (Mb)\\
\hline\\[-0.3cm]
 resonance 1s$\to$3p [configuration 1s2s$^2$2p$^6$3p $^1$P]\\[0.1cm]
  TDHF                       & 888.1  &  2.420 & 653.3  & 0.01815  & 0.120  &  0  &  914  \\
  TDRSH                      & 832.3  &  1.432 & 203.8  & 0.01790  & 0.552  &  0  &  857 \\
  TDLRSH-K                   & 864.9  &  2.237 & 347.7  & 0.01648  & 0.289  &  0  &  890 \\
  TDLRSH-AK                  & 867.0  &  2.225 & 370    & 0.01640  & 0.265  &  0  &  594 \\
  EOM-CCSD$^\text{a}$        & 866.9\\
  RCI$^\text{b}$             & 865.8 \\
  Experiment$^\text{c}$      & 867.3 & 248(2)\\[0.2cm]

 resonance 1s$\to$4p [configuration 1s2s$^2$2p$^6$4p $^1$P]\\[0.1cm]
  TDHF                       & 890.2  &  0.465 & 550.6  & 0.01900 & 0.162  &  0  &  935  \\
  TDRSH                      & 833.9  &  0.381 & 199.5  & 0.01807  & 0.571  &  0  &  410 \\
  TDLRSH-K                   & 866.7  &  0.524 & 321.0  & 0.01760 & 0.318 &  0  &  576 \\
  TDLRSH-AK                  & 868.8  &  0.518 & 339.3  & 0.01747  & 0.296  &  0  &  595 \\
  RCI$^\text{b}$             & 867.3\\
  Experiment$^\text{c}$      & 868.9 & 260(3)\\[0.2cm]

 resonance 2s$\to$3p [configuration 1s$^2$2s2p$^6$3p $^1$P]\\[0.1cm]
  TDHF                          &  49.578 & 22.797 & -0.932 & 8.765  & 0.872 &  0 & 7.8 \\
  TDRSH                         &  43.415 & 23.768 & -1.419 & 8.859  & 0.610 &  0 & 14.3 \\
  TDLRSH-K                      &  43.664 & 23.877 & -1.414 & 8.886  & 0.607 &  0 & 14.3 \\  
  TDLRSH-AK                     &  43.829 & 24.084 & -1.388 & 8.892  & 0.615 &  0 & 13.8 \\  
  R-matrix$^\text{d}$           & 45.534 & 34.9 \\
  XCI$^\text{e}$                & 45.431 & 15.0 & -1.47 & & 0.79 &  & \\
  RCI$^\text{b}$                & 46.2\\
  Experiment$^\text{f}$         & 45.546(8)  & 13(2)    & -1.6(2) &   &    0.70(7) &    &  \\
  Experiment$^\text{d}$         & 45.544(5) & 16(2)\\[0.2cm]

  resonance 2s$\to$4p [configuration 1s$^2$2s2p$^6$4p $^1$P]\\[0.1cm]
  TDHF                        &  51.133 & 5.912 & -1.038 & 8.602  & 0.883 &  6.8$\cdot 10^{-5}$ & 9.2 \\
  TDRSH                       &  44.891 & 6.531 & -1.508 & 8.720  & 0.619 &  9.6$\cdot 10^{-5}$ & 15.6  \\
  TDLRSH-K                    &  45.144 & 6.545 & -1.503 & 8.747  & 0.616 &  9.5$\cdot 10^{-5}$ & 15.5  \\  
  TDLRSH-AK                   &  45.310 & 6.593 & -1.480 & 8.509  & 0.621 &  8.5$\cdot 10^{-5}$ & 15.2  \\
  R-matrix$^\text{d}$         & 47.111 & 6.65 \\
  XCI$^\text{e}$              & 46.942 & 4.3 & -1.26 & & 0.84 &  & \\
  RCI$^\text{b}$              & 47.6\\
  Experiment$^\text{f}$       & 47.121(5)  & 5(2)    & -1.6(3) &   &    0.70(7) &    &  \\
  Experiment$^\text{d}$       & 47.119(5) & 6.6\\
  \hline\hline
\multicolumn{5}{l}{$^\text{a}$From Ref.~\onlinecite{SkoKry-JCP-21b}.}\\
\multicolumn{5}{l}{$^\text{b}$From Ref.~\onlinecite{AlnHamAbu-CPI-25}.}\\
\multicolumn{5}{l}{$^\text{c}$From Ref.~\onlinecite{MulBerBorBuhHelHolKilKluMarRicSelVieChi-AJ-17}.}\\
\multicolumn{5}{l}{$^\text{d}$From Ref.~\onlinecite{SchDomPutGutKaiMieGre-PRA-96}.}\\
\multicolumn{5}{l}{$^\text{e}$From Ref.~\onlinecite{MarKliKjeLinGonArgMar-PRA-17}.}\\
\multicolumn{5}{l}{$^\text{f}$From Ref.~\onlinecite{CodMadEde-PR-67}.}
\end{tabular}
\end{table*}

For the two valence resonances, TDHF gives overestimated resonance energies, while TDRSH and TDLRSH give underestimated resonance energies. The resonance energies given by TDLRSH-AK, 43.829 and 45.310 eV, are the closest to the experimental values, 45.544 and 47.119 eV, for the 2s $\to$ 3p and 2s $\to$ 4p resonances, respectively. The errors (1.7 and 1.8 eV) are still substantial and must be related to the fact that none of the parameters of TDLRSH-AK have been optimized on the 2s ionization edge. In fact, these errors are about the same as the error in the TDLRSH-AK 2s orbital energy compared to the exact 2s ionization edge (1.8 eV), displayed in Figure~\ref{fig:orbitalenergiescHD}, which is not a surprise since the  2s $\to$ 3p and 2s $\to$ 4p resonances are close to the 2s ionization edge. By contrast to the situation for core resonances, TDHF, TDRSH, and TDLRSH all give valence resonance widths with correct order of magnitude. In particular, TDLRSH-AK gives widths of 24 and 6.6 meV for the 2s $\to$ 3p and 2s $\to$ 4p resonances, respectively, while the more recent experimental values are 16 and 6.6 meV. TDRSH and TDLRSH also give Fano asymmetric parameters $q$ in good agreement with values from experiment or accurate calculations. In the Ne atom, the 2s $\to$ $n$p resonances autoionize via a participator Auger-like process~\cite{BurTay-JPB-75,SchDomPutGutKaiMieGre-PRA-96,LanBerWehGorBozFar-JPB-97}
\begin{equation}
1\text{s}^2 2\text{s} \, 2\text{p}^6 n\text{p} \to 1\text{s}^2 2\text{s}^2 2\text{p}^5  + \text{e}^-. 
\end{equation}
Since the states $1\text{s}^2 2\text{s}^2 2\text{p}^5$ are single excitations with respect to the ground-state configuration, they are explicitly taken into account in linear-response TDHF or adiabatic single-determinant TDDFT. This explains why TDHF, TDRSH, and TDLRSH reasonably describe the lifetimes of these resonances.

\section{Conclusion}
\label{sec:conclusion}

In this work, we have tested range-separated hybrids and locally range-separated hybrids for linear-response TDDFT Sternheimer calculations of the photoionization spectrum of the Ne atom. Obtaining an accurate photoionization spectrum is challenging for TDDFT with the usual exchange-correlation approximations since it involves core and valence resonances with very different energy scales. We have shown that the locally range-separated hybrid using the two-parameter range-separation function in Eq.~\eqref{murAK} quite satisfactorily describes the photoionization spectrum overall, including the energies of both core and valence resonances. This is achieved thanks to a proper position-dependent removal of the self-interaction error. The lifetimes of the 2s $\to$ $n$p valence resonances are reasonably described as well, owing to the fact that their decay does not involve double excitations. However, the lifetimes of the 1s $\to$ $n$p core resonances are overestimated by two orders of magnitudes or more, which can be explained by the fact that their decay via Auger processes involving double excitations, which are not explicitly taken into account within adiabatic single-determinant TDDFT. 

In light of the present results, it seems clear that, to obtain more accurate resonance widths with linear-response range-separated TDDFT, the effect of double excitations should be included. This could be achieved by using linear-response multideterminant range-separated TDDFT~\cite{FroKneJen-JCP-13} and/or by adding a frequency-dependent response kernel~\cite{RebTou-JCP-16}.


\section*{Author Declarations}
The authors have no conflicts to disclose.

\section*{Data Availability}
The data that support the findings of this study are available from the corresponding author upon reasonable request.



\end{document}